\documentclass[letterpaper, 10 pt, conference]{cssconf}  

\IEEEoverridecommandlockouts                              

\usepackage{amssymb, latexsym, epsfig}

\usepackage{tabularx}

\usepackage{color}

\title{\LARGE \bf Robustness in Gene Circuits: Clustering of Functional Responses}

\author{Mary J. Dunlop and Michael E. Wall
\thanks{M. J. Dunlop is in the Division of Engineering and Applied Science,
        California Institute of Technology, CA 91125, USA.
        {\tt\small mjdunlop@caltech.edu}}%
\thanks{M. E. Wall is in the Computer and Computational Sciences Division and Bioscience Division, Los
	Alamos National Laboratory,
        Los Alamos, NM 87545, USA.
        {\tt\small mewall@lanl.gov}}%
}

\begin{document}

\maketitle
\thispagestyle{empty}
\pagestyle{empty}

\begin{abstract}

In contrast to engineering applications,
in which the structure of control laws are designed to satisfy
prescribed function requirements, in biology it is often necessary to
infer gene-circuit function from incomplete data on gene-circuit structure. By
using the feed-forward loop as a model system, this paper introduces a technique for
classifying gene-circuit function given gene-circuit structure. In
simulations performed on a comprehensive set of models that span a broad
range of parameter space, some designs are robust, producing one unique
type of functional response regardless of parameter selection. Other
designs may exhibit a variety of functional responses, depending upon
parameter values. We conclude that, although some feed-forward loop models have
designs that lend themselves to unique function inference, others have
designs for which the function type may be uncertain. 

\end{abstract}

\section{INTRODUCTION}

\begin{figure}[thpb]
      \centering
      \includegraphics[scale=0.65]{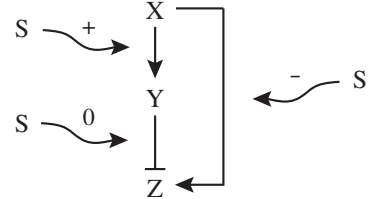}
      \caption{Feed-forward Loop Network Motif. $X$ and $Y$ represent transcription factors. 
	       $Z$ is the target (effector) gene.
	       Activator connections are drawn as normal arrows and repressor connections are drawn as arrows with
	       T-shaped ends.
	       Signal effects are shown with characters \{+, -, 0\}.}
      \label{ffl}
   \end{figure}

In cells, gene expression is often influenced by molecular signals. The genes and gene products involved in the
response to a signal comprise a genetic regulatory circuit, and the set of genetic regulatory interactions in a 
cell defines its genetic regulatory network. As with 
engineered systems, there are certain designs that may recur in different parts of a cell's genetic 
regulatory network. For example, it has recently been found that certain patterns of genetic regulatory interactions
occur more frequently in {\it Escherichia coli} than would be expected in 
randomized networks with similar connection statistics \cite{AlonNature2002}. The feed-forward loop is one such design, 
an example being regulation of {\em araBAD} by both the local transcription factor AraC and the global transcription 
factor CRP in {\it E. coli} (see review \cite{Scheif2000} and references therein). 
An example of a feed-forward loop with an arbitrarily selected configuration is shown in Fig. \ref{ffl}.

Given that there are recurring structural designs found in genetic regulatory networks, it is logical to ask: {\it a)} What
is the function of a design and {\it b)} why might one design be preferred over other designs?  

The first question addresses the issue that even once a particular circuit 
configuration is selected, the function of the circuit is not necessarily transparent. For the feed-forward loop, 
Mangan and Alon \cite{ManganAlonPNAS2003} explored several possible circuit functions by using a mathematical model
of feed-forward loops in which a signal $Sx$ interacts with $X$, and a different signal $Sy$ interacts with $Y$. For
a constant level of $Sy$, they noticed pulsing, on/off, and off/on behaviors in gene expression levels in response to a step
function input in $Sx$. Their efforts produced a preliminary classification of functional responses for feed-forward
loops, but in order to more thoroughly characterize feed-forward loop function, it is desirable to explore a larger
range of parameters than were considered in their study, and to consider circuit types in which the same signal can
interact with both $X$ and $Y$.

Answering the second question requires an understanding of performance criteria relevant to natural selection in gene
circuits. Elucidation of design principles is a subject of interest to biologists and engineers alike
\cite{WallNature2004}. Broad classification of possible circuit functions can eventually help clarify why certain
circuit designs are preferable to others. For example, the feed-forward loop simulations presented in this paper show
that the function of some circuit designs is more robust to parameter changes than for other circuit designs. 

This paper presents a method for classifying possible functions of a gene-circuit given the structural 
configuration. Temporal responses of a comprehensive set of feed-forward loop models are calculated for a range of
parameter values. The responses are clustered, and the relation between clusters and circuit types is analyzed.

\section{FEED-FORWARD LOOP} \label{fflsection}

Protein levels in bacteria are often controlled at the level of transcription. Proteins called transcription factors can 
bind to regulatory regions of DNA upstream of a gene of interest to either assist in enabling transcription or 
to block the transcription process. When transcription occurs, the downstream genetic information is copied to 
a strand of mRNA and then translated into a protein, which can effect a physical change in the cell (e.g. break 
down sugars present in the environment). The protein encoded by the expressed gene could, alternatively, be another 
transcription factor that will activate or repress the transcription of a gene elsewhere in the cell. Because of this structure, it is possible to see chains of transcription factors that ultimately lead to the expression of a 
target gene. 

The feed-forward loop network motif has 2 transcription factors, $X$ and $Y$, which control expression levels of a
target gene $Z$. $X$ additionally regulates transcription of $Y$. The term {\it functional response} refers to the expression level of $Z$ as a function of time. 

If the presence of a transcription factor $X$ enables transcription of $Y$, then it is said that 
$X$ {\it activates} $Y$. If the presence of $X$ inhibits transcription of $Y$, $X$ {\it represses} $Y$. 

Signaling molecules also play a significant role in gene expression. Signals may be small-molecule metabolites or other
molecules that bind to the transcription factor, enabling or blocking its activity. 

As in Mangan \& Alon \cite{ManganAlonPNAS2003}, we consider feed-forward loop models in which each of 3 genetic regulatory interactions can take on one of two possible values (\{activator, 
repressor\}). Unlike their study, which only considers changes in a signal that enables the global activity of $X$,
we consider models in which a signal may have one of 3 effects (\{+, -, 0\}) on each genetic regulatory interaction.
Instead of considering just 8 (= $2^3$), we consider 216 (= $2^33^3$) different ways of wiring a feed-forward loop. 
Fig. \ref{ffl} is just one example.

\section{MATHEMATICAL MODELS} \label{math_models}

The general feed-forward loop is modeled using a pair of nonlinear ordinary differential equations: 

\begin{eqnarray} 
{\dot Y} &=& B_y + \alpha_y H\left({S_{yx}X \over K_{yx}}\right) - \beta_y Y \label{odeY} \\
{\dot Z} &=& B_z + \alpha_z H\left({S_{zy}Y \over K_{zy}}\right) H\left({S_{zx}X \over K_{zx}}\right) - \beta_z Z. \label{odeZ}
\end{eqnarray}

\noindent Recall, $X$ and $Y$ are transcription factors and $Z$ is the target gene. $X$ is treated as an independent
variable, modeled here as a constant, as in \cite{ManganAlonPNAS2003}. This assumes constitutive production of $X$;
formation of $X$ does not depend on the presence of any specific substrate.

$B_i$ is the basal transcription rate, the rate of protein production that cannot be controlled through 
transcriptional regulation. $\alpha_i$ is the regulatable transcription rate and $\beta_i$ is the decay 
rate through degradation and dilution. $S_{ij}$, discussed in further detail below, is a binary value that describes the signal effect. $K_{ij}$ is a threshold value, also described below. 

The Hill function, $H(x)$, that appears in Eqns. \ref{odeY}--\ref{odeZ} describes how well the transcription factor 
is bound to the DNA. It maps the ratio of a transcription factor level to threshold (e.g. $X/K_{yx}$) to a scalar value between $0$ 
and $1$ that describes how effectively the transcription process occurs. High values indicate more effective
transcription. 

The Hill function is described by

\begin{eqnarray}
H(x) = {1 \over (1 + x^{n_{ij}})}, \label{hillfunctioneqn}
\end{eqnarray}

\noindent where $n_{ij}$ is negative if the connection is an activator and positive if it is a repressor. Reference
\cite{Murray_mathbio}
provides a general introduction to equations like \ref{odeY}--\ref{hillfunctioneqn}. 

The Hill function coefficient $n_{ij}$ determines how rapidly the function transitions between 0 and 1. As $|n_{ij}| \rightarrow \infty$, the Hill function becomes a step function. Real biological systems tend to have $|n_{ij}|$ close to 2. Step function approximations can be solved analytically, however.

The threshold value $K_{ij}$ is the value of $j$ at which the Hill function is equal to 0.5. The Hill function can be derived 
from equations describing the chemistry of DNA/transcription factor binding.  



Signal interactions are modeled by inserting a binary term, $S_{ij} \in \{0, 1\}$, in the Hill function argument. 
The value $S_{ij}$ takes on depends upon the level of signal in the environment and the type of signal 
interaction (\{+, -, 0\}). Table \ref{Sijvals} is used to determine $S_{ij}$.

\begin{table}[h]
\begin{center}
\begin{tabular} {|c|c|c|}
\hline
& Signal $<$ Threshold & Signal $>$ Threshold\\ \hline
+ & 0 & 1\\ \hline
- & 1 & 0\\ \hline
0 & 1 & 1\\ \hline
\end{tabular}
\caption{$S_{ij}$ Values} \label{Sijvals}
\end{center}
\end{table}

\section{SIMULATIONS} \label{simulations}

The initial conditions for all simulations are the steady state values of $Y$ and $Z$ when the signal is below
the threshold level. For the feed-forward loop, analytical expressions for $Y(0)$ and $Z(0)$ can be found from 
the steady state versions of Eqns. \ref{odeY}
and \ref{odeZ}. We are interested in the dynamical behavior that results from changing signal levels. 

\begin{figure}[thpb]
      \centering
      \includegraphics[scale=0.62]{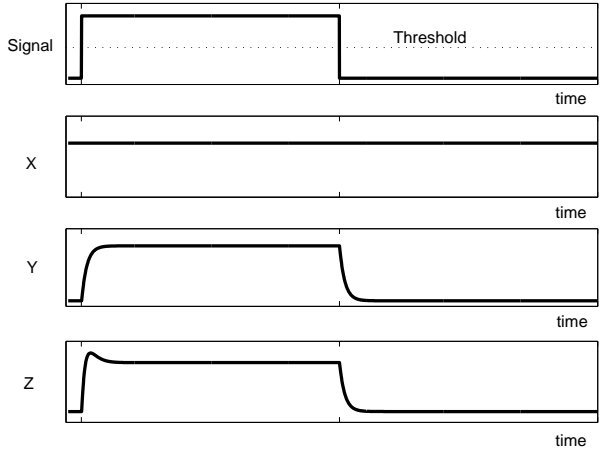}
      \caption{Sample functional response. Simulation results for levels of $X$, $Y$, and $Z$ expression as a 
               function of time. The input signal and threshold are shown in the top plot.
		The conf{i}guration of the simulated system is \{X-Y, Y-Z, X-Z\} = \{activator, repressor, activator\} and 
		\{Signal X-Y, Signal Y-Z, Signal X-Z\} = \{+, +, +\}. 
		Parameter values shown here are: $\alpha_i = \beta_i = K_{ij} = 1$, $B_i = 0.1$, $X = 10$, $|n_{ij}| = 2$.}
      \label{sample_func_response}
   \end{figure}

Fig. \ref{sample_func_response} shows the response of one representative feed-forward loop to changing signal levels. 
The level of transcription factor $Y$, increases to its steady state value following a decaying exponential curve. Nonlinear
effects cause overshoot in $Z$ before it reaches steady state.  

There are 9 parameters in Eqns. \ref{odeY}--\ref{odeZ} that can be varied. 

Through a change of variables $B_i$ ($i~=~y,~z$) can be eliminated from the equations. 
Its presence causes multiplicative and additive shifts in the functional response. 
As discussed in Section \ref{svdsection}, neither of these 
properties are important in clustering. The $B_i$ parameters are arbitrarily set to zero.

The values of $\alpha_i$ and $\beta_i$ are selected randomly from a reasonable range ([0.1, 10] for data shown 
in the following section). For each of the 4 parameters, a random number, $r$, is mapped using a power law:
                                                                                                  
\begin{eqnarray}
10^{(2r-1)log_{10}M_{\alpha \beta}},
\end{eqnarray}
                                                                                                  
\noindent where $M_{\alpha \beta}$ is the maximum value $\alpha_i$ or $\beta_i$ can take on (10 for this example). 
This mapping ensures that it is equally likely to assign values less and greater than 1 to the parameters.
                                                                                                  
Additionally, the threshold parameters, $K_{ij}$, are varied. The ratio of transcription factor concentration to
the threshold value is the relevant quantity (e.g. ${X \over K_{yx}}$). These three ratios are allowed to
take on values less than 1, equal to 1, and greater than 1. All 27 possible combinations are considered.

Recognizing symmetry in signaling effects reduces the size of this problem and decreases computation time. 

\section{CLUSTERING FUNCTIONAL RESPONSES} \label{clustering}

An infinitely large number of feed-forward loops can be modeled with these techniques. For each of the 216 
wiring patterns there are multiple threshold and rate parameters that are either unknown or uncertain in 
biological systems. Broad range limits on parameter values can be assumed to make the problem tractable, but
the number of systems remains large.

Although a great number of feed-forward loops can be modeled through this approach, many of the functional 
responses appear to be similar. The number of systems we consider is so large that an automated approach is 
necessary to classify functional responses into different categories based upon their similarity. A 
clustering algorithm is used to tackle this problem.

\subsection{Clustering Algorithm}
The method used to cluster functional responses is a greedy approximation algorithm developed by Hochbaum
and Shmoys \cite{HochbaumShmoys1985}.
The algorithm requires a metric $d(x, y)$ on a set $Y$ ($x, y \in Y$) that 
characterizes the distance between $x$ and $y$. For a point $y \in Y$ and $S \subseteq Y$ define
$d(y, S) = \min \{d(y, s) : s \in S\}$. 

The clustering problem this algorithm solves can be stated as follows: 
Given an input of a set $X$ of $n$ points $x_1,~...~x_n$ and a metric $d$ on $X$, find a set $C$ of $K$ points 
$c_1,~...~c_K \in X$ that minimizes $\max \limits_{1 \leq i \leq n} d(x_i, C)$.


The clustering algorithm is performed as follows. First, $K$ points must be selected as cluster centers. The
first center, $c_1$, is chosen at random. After that ($i = 2,~...~K$) let $c_i$ be the point $x$ of $X$ 
that maximizes $d(x, \{c_1,~...~c_{i-1}\})$. This is equivalent to assigning all the remaining non-center 
points to clusters, determining which is furthest from its center point, assigning that point as 
a new center, and throwing the rest of the points back into the pool of non-centers. 
After all $K$ centers have been assigned, the remaining points $x_{K+1},~...~x_n$ are assigned to clusters.

This algorithm is used to cluster functional responses. Defining a distance measure, $d$, 
is the primary complication in extending the clustering algorithm to the present task. Each functional 
response is a vector $\underline{z} \in \mathbb{R}^N$ where the vector contains the values of $Z$ running from
$t = 0$ to $t = N-1$.

A correlation coefficient is used to measure the distance between two functional response vectors, 
$\underline{z}_1$ and $\underline{z}_2$: 

\begin{eqnarray}
d(\underline{z}_1, \underline{z}_2) = 
{1 \over 2} - 
{
< \underline{z}_1 - \bar{\underline{z}_1}~,~\underline{z}_2 - \bar{\underline{z}_2} >
\over
2~~||\underline{z}_1 - \bar{\underline{z}_1} ||_2~~|| \underline{z}_2 - \bar{\underline{z}_2} ||_2
}.
\label{corrcoeff}
\end{eqnarray}

\noindent This distance function is designed so that $d(\underline{z}_1, \underline{z}_2) = 0$ if $\underline{z}_1 = \underline{z}_2$ and $d(\underline{z}_1, \underline{z}_2) = 1$ if the two signals are very different. 

\subsection{Maximum Error versus Number of Clusters}
The maximum error is defined as the largest cluster ``radius", $\max \limits_{1 \leq i \leq n} d(x_i, C)$. 
This value can be plotted as a function of $K$, the number of clusters. 
Clearly, at $K$ = $1$, we will have a large maximum error value if the functional responses are not all 
nearly identical. For $K$ = total \# of functional responses, we will have 
$0$ error since every functional response is associated with its own individual cluster. The shape of the 
intermediate curve is of particular interest. 

   \begin{figure}[thpb]
      \centering
      \includegraphics[scale=0.6]{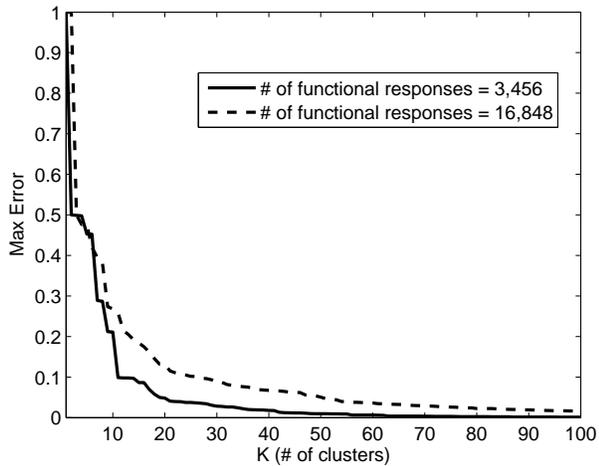}
      \caption{Maximum Error versus $K$, the number of clusters. The results of clustering on two distinct data sets are shown.}
      \label{kvserror}
   \end{figure}

Fig. \ref{kvserror} shows that, even for large numbers of functional
responses, the maximum error drops off rapidly as the number of clusters is 
increased.  

\subsection{Singular Value Decomposition} \label{svdsection}
Since each cluster may contain a large number of functional responses, plotting 
all the functional responses associated with an individual cluster on top of one another is overwhelming.
In some clusters, there may be thousands of responses. Singular value decomposition 
is used to generate a representative trace that describes the most significant principal component of all of the 
functional responses in the cluster.

Singular value decomposition has been used in other biological applications to compress data into a 
simplified, more understandable form \cite{SVDChapter}. In this work the singular value decomposition of
a matrix $A \in \mathbb{R}^{M \times N}$ is taken: 

\begin{eqnarray}
A = U S V^T. \label{svd}
\end{eqnarray}

\noindent $M$ is the number of functional responses we are comparing and $N$ is the number points in time. 
S, U, and V come from the standard definition of singular value decomposition.

The first right singular vector (the first column of $V$) is the singular vector associated with the largest
singular value. This vector describes the principal component of all of the functional responses
listed in the $A$ matrix and provides a single representative functional response to associate with a cluster. 

   \begin{figure}[thpb]
      \centering
      \includegraphics[scale=0.6]{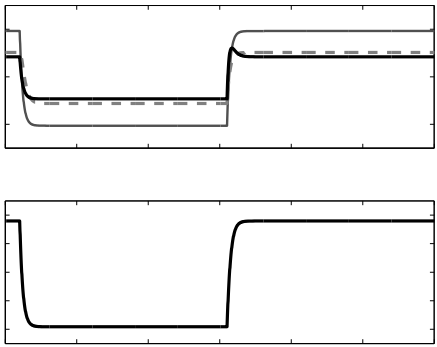}
      \caption{Singular Value Decomposition Example. a) Three representative functional responses from one cluster. 
		b) The f{ir}st right singular vector of a matrix containing all functional responses from the cluster that the
responses shown in a) are drawn from.}
      \label{svdexample}
   \end{figure}

An example of how singular value decomposition can be used to represent many functional responses is shown in 
Fig. \ref{svdexample}. The top plot of Fig. \ref{svdexample} shows 3 functional responses plotted on top of each other. In 
reality, this is a small subset of all functional responses that fall into this cluster type. The primary 
singular vector associated with the complete set of functional responses is shown in the bottom plot of Fig. \ref{svdexample}. 
Note that the distance function (Eqn. \ref{corrcoeff}) evaluates to zero for functional responses that differ 
only by a multiplicative scaling factor and an offset. 

\section{Results} \label{results}

The clustering approach associates the functional responses of 216 feed-forward loop models with a small number of
distinct patterns. These distinct patterns can be used to classify the behavior of an
individual circuit over a range of parameter values. The number of clusters it takes to describe a particular 
circuit configuration can be used as a measure of how robust a circuit is to parameter variation.

\subsection{Representative Cluster Traces}
\begin{figure}[thpb]
      \centering
      \includegraphics[scale=0.75]{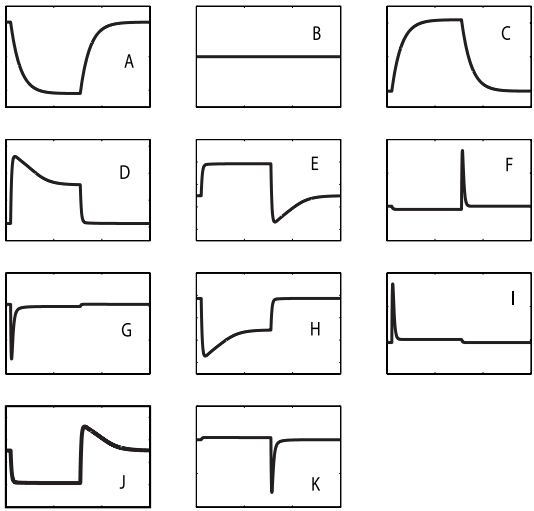}
      \caption{Representative functional responses ($Z$ vs. $time$) from each of 11 clusters. 
		Letter labels are used for reference in Table \ref{clustertable}.}
      \label{clusters11}
   \end{figure}

A relatively simple example is presented to illustrate the utility of clustering. The data shown in 
Fig. \ref{clusters11} are the result of clustering on a set of 3,456 functional responses. All 216 possible 
circuit configurations are represented. Within each configuration only parameters $\alpha_i$ and $\beta_i$ 
($i~=~y,~z$) from Eqns. \ref{odeY}--\ref{odeZ} are varied. 

K = 11 clusters is chosen as a cutoff point because the maximum error is acceptably small (see \# of functional 
responses = 3,456 in Fig. \ref{kvserror}). Beyond this point additional cluster types represent similar functional 
responses but with differing temporal characteristics. For example, the rise times, settling times, and overshoot 
behavior may be different for the additional cluster types. The utility
of clustering lies in its ability to segregate functional responses into broad class types, 
allowing for a qualitative understanding of possible circuit functionality. In particular, this method will be useful for 
considering circuits that have more complicated responses (e.g. responses to input signals that are more complicated
than a step function).

\begin{figure}[thpb]
      \centering
      \includegraphics[scale=0.2]{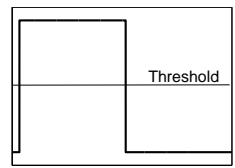}
      \caption{Signal level as a function of time.}
      \label{signaltrace}
   \end{figure}

Fig. \ref{clusters11} shows representative singular vectors from each of the 11 clusters. These are the functional 
responses ($Z$ vs. $time$) of various systems in response to the input signal shown in Fig. \ref{signaltrace}. 

\begin{table*}[thpb]
\begin{center}
\begin{tabular} {|@{}c@{}c@{}c@{}c@{}c@{}c|p{4mm}p{4mm}p{4mm}p{4mm}p{4mm}p{4mm}p{4mm}p{4mm}p{4mm}p{4mm}p{4mm}p{4mm}p{4mm}p{4mm}p{4mm}|p{8mm}|}
\hline
 & & & Signal & Signal & Signal & & & & & & & & & & & & & & & &\\
X-Y & Y-Z & X-Z & X-Y & Y-Z & X-Z & A & B & C & D & E & F & G & H & I & J & K & L & M & N & O & Entropy\\ 
\hline
act & rep & rep & + & + & + & 1.00 &    0 &    0 &    0 &    0 & 0 & 0 & 0 &    0 & 0 &    0 & 0 &    0 &    0 & 0 & 0\\	   
act & rep & rep & + & + & 0 & 1.00 &    0 &    0 &    0 &    0 & 0 & 0 & 0 &    0 & 0 &    0 & 0 &    0 &    0 & 0 & 0\\	   
act & rep & rep & + & + & - & 0.41 &    0 & 0.51 & 0.02 &    0 & 0 & 0 & 0 & 0.01 & 0 &    0 & 0 &    0 & 0.06 & 0 & 1.45\\	   
act & rep & rep & + & 0 & + & 1.00 &    0 &    0 &    0 &    0 & 0 & 0 & 0 &    0 & 0 &    0 & 0 &    0 &    0 & 0 & 0\\	   
act & rep & rep & + & 0 & 0 & 1.00 &    0 &    0 &    0 &    0 & 0 & 0 & 0 &    0 & 0 &    0 & 0 &    0 &    0 & 0 & 0\\	   
act & rep & rep & + & 0 & - & 0.38 &    0 & 0.51 & 0.03 &    0 & 0 & 0 & 0 & 0.01 & 0 &    0 & 0 &    0 & 0.07 & 0 & 1.51\\	   
act & rep & rep & + & - & + & 0.90 &    0 &    0 &    0 & 0.09 & 0 & 0 & 0 &    0 & 0 &    0 & 0 & 0.01 &    0 & 0 & 0.52\\	   
act & rep & rep & + & - & 0 &    0 & 0.11 &    0 &    0 & 0.06 & 0 & 0 & 0 &    0 & 0 & 0.50 & 0 & 0.33 &    0 & 0 & 1.62\\	   
act & rep & rep & + & - & - &    0 &    0 & 1.00 &    0 &    0 & 0 & 0 & 0 &    0 & 0 &    0 & 0 &    0 &    0 & 0 & 0\\	   
act & rep & rep & 0 & + & + & 1.00 &    0 &    0 &    0 &    0 & 0 & 0 & 0 &    0 & 0 &    0 & 0 &    0 &    0 & 0 & 0\\	   
act & rep & rep & 0 & + & 0 & 1.00 &    0 &    0 &    0 &    0 & 0 & 0 & 0 &    0 & 0 &    0 & 0 &    0 &    0 & 0 & 0\\	   
act & rep & rep & 0 & + & - & 0.48 &    0 & 0.52 &    0 &    0 & 0 & 0 & 0 &    0 & 0 &    0 & 0 &    0 &    0 & 0 & 1.00\\	   
\hline
\end{tabular}
\caption{Percentage of Cluster Types Exhibited By Circuit Configurations \break (Subsection of a Larger Table)} 
\label{clustertable}
\end{center}
\end{table*}

When the threshold values, $K_{ij}$, are varied in addition to $\alpha_i$ and $\beta_i$ all relevant parameters in Eqns. \ref{odeY}--\ref{odeZ} are explored. 
The result is a large set of functional
responses that do not segregate as logically into individual clusters. Even when this is the case, the clustering 
technique can still be applied to yield a qualitative picture of possible functional responses. The case with 16,848 functional
responses shown in Fig. \ref{kvserror} corresponds to a widely explored range of parameters, but the maximum error still
drops off rapidly. If an acceptable error value is chosen, clustering can be performed to within this margin
of error.

\begin{figure}[thpb]
      \centering
      \includegraphics[scale=0.75]{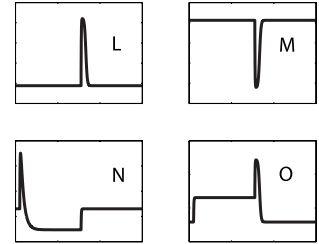}
      \caption{Representative functional responses ($Z$ vs. $time$) for cluster types added when 
		additional parameters are varied. 
		Letter labels are used for reference in Table \ref{clustertable}.}
      \label{clusters_plus4}
   \end{figure}

In an exploration of the more complete parameter space, the cluster types seen in Fig. \ref{clusters11} 
are preserved, but several additional clusters are added.
For example, selecting K = 15 clusters with 16,848 functional responses produces the 4 additional cluster types shown in Fig. \ref{clusters_plus4}.


Even when exploring the complete parameter space, some system configurations fall into the same cluster type 
regardless of the parameter values selected for $\alpha_i$, $\beta_i$, and $K_{ij}$. The functional responses of
these genetic regulatory 
configurations are robust because they are particularly insensitive to changes in parameters.  

\subsection{Distributions of Functional Responses}

Table \ref{clustertable} lists cluster types for various system configurations. Each row corresponds to one 
particular configuration, a set of genetic regulatory and signal interaction types ({\it act} = activator, {\it rep} = repressor). The columns labeled A--O correspond to the cluster types labeled in Figs. \ref{clusters11} and 
\ref{clusters_plus4}. The numbers in the row tell into which cluster this system's functional responses fall. 
For some system configurations, varying parameter values causes the functional response to fall 
into different clusters. The rows are 
normalized by the total number of cases with different parameter values that were run. 

Table \ref{clustertable}'s entries list the percentages of functional responses that fall into each
cluster type for each configuration, but these data do not indicate how ``different" responses are within a cluster. Selecting the
number of clusters (Fig. \ref{kvserror}) sets the upper bound on the error within each cluster. For the 15
cluster case, all functional responses within a cluster are within a distance of 0.18 of each other, as measured
by Eqn. \ref{corrcoeff}.

The entropy of each function distribution in Table \ref{clustertable} is calculated by using the standard definition 
of Shannon entropy:

\begin{eqnarray}
- \sum _{i=1} ^{15} p(i)~log_2~p(i), \label{entropy}
\end{eqnarray}

\noindent where $p(i)$ is the percentage of functional responses that fall into cluster $i$. 

In the context of the previous discussion on robustness, rows which have a 1 associated with one cluster type 
and 0's for all the rest (entropy = 0) are particularly robust because parameter variations do not change the 
function of the system. 

Note that the exact entries in the table are dependent upon the details of parameter selection and the distance measure
used to cluster data. General trends, however, can still be concluded from these data.

The entries shown in Table \ref{clustertable} are a subset from a larger table. If all possible genetic
regulatory and signal interaction types were shown, the table would have 216 rows. The 12 configurations types shown are
illustrative examples; although these circuits do not exhibit a significant number of responses that fall within
clusters $F$, $G$, $H$, $L$, and $O$, other circuits may exhibit such responses, and have patterns of diversity
similar to those shown here.

\subsection{Biological Interpretation}
Clustering provides a logical grouping of response types without prior knowledge of the behavior that a network
may exhibit. The response pattern of a cluster can then be interpreted in a biologically meaningful way. For example,
cluster $A$ (Fig. \ref{clusters11}) is associated with repressible circuits, for which gene expression decreases upon 
an increase in signal level. Cluster $C$ is similarly associated with inducible circuits, for which gene expression
increases upon increase in signal. For circuits associated with cluster $B$, gene expression is unresponsive to
changes in signal. Pulsed gene expression responses, both with and without steady state changes, are seen in the remaining cluster types.

\section{CONCLUSIONS AND FUTURE WORK}


This paper presents a method for identifying functional capabilities of a genetic network given its structure.
The feed-forward loop network motif is used to demonstrate the utility of the technique. 
In our analysis of feed-forward loop models, functional responses were organized into a relatively small number of
clusters. Some feed-forward loop types show non-robust behavior, suggesting that these circuits do not
have unique information processing roles. The clustering
technique presented allows for such quick, qualitative intuition into the function of a 
system. Insight from clustering will be particularly helpful if the state space and parameter space are even larger
than those presented in the feed-forward loop example here. 

Although we consider models of feed-forward loops in isolation, in Nature gene circuits are embedded within the
context of the entire molecular network of the cell. Nevertheless, considering isolated gene-circuit models can
reveal insights into the cellular response to signals. Such models have already proved to be useful in design
of synthetic gene circuits, for example, in the design of a toggle
switch \cite{Gardner2000}, an oscillator \cite{Elowitz2000}, and a circuit whose design may be selected to exhibit
either toggle switch or oscillatory behavior \cite{Atkinson2003}.
The present technique can help to narrow down which system types and parameter ranges exhibit a desirable
behavior, given a broad class of possible designs.

In the future it will be interesting to explore the implications of robustness of functional responses in real biological systems. In particular, is
robustness necessarily a desirable trait for a genetic circuit? If the circuit is locked into one role it may not be capable of
evolving in alternative environments. However, robustness leads to reliable functionality. In addition, natural
selection can act to enhance the populations of organisms that are sensitive rather than robust to mutations in gene
circuits. This process has been used previously to explain patterns in the use of activator and repressor
control in natural genetic regulatory interactions \cite{Savageau1977}. It would be interesting to consider tradeoffs
involving robustness in the context of the evolution of feed-forward loop configurations and other aspects of gene
circuit design.

\section{ACKNOWLEDGMENTS}

We thank Richard Murray for his critical reading of the manuscript. 

This work was supported by the Department of Energy both under
contract W-7405-ENG-36 to the University of California (M.E.W.), and
through the DOE Computational Science Graduate Fellowship Program (M.J.D.).

\bibliography{references}

\end{document}